\documentclass[iop]{emulateapj}

\newcommand{\icarus}{Icarus}

\newcommand{\msun}{M$_{\odot}$}
\newcommand{\lsun}{L$_{\odot}$}
\newcommand{\mjup}{M$_{\mathrm{Jup}}$}
\newcommand{\mearth}{M$_{\mathrm{\oplus}}$}

\slugcomment{To appear in ApJ}

\shorttitle{Remnant gas in evolved circumstellar disks}
\shortauthors{Geers et al.}

\begin{document}

\title{Remnant gas in evolved circumstellar disks: \\Herschel PACS observations of 10--100 Myr old disk systems\altaffilmark{*}}

\author{Vincent C. Geers\altaffilmark{1}, Uma Gorti\altaffilmark{2,3}, Michael R. Meyer\altaffilmark{1}, Eric Mamajek\altaffilmark{4,5}, Arnold O. Benz\altaffilmark{1}, David Hollenbach\altaffilmark{3}}

\email{Electronic address: vcgeers@phys.ethz.ch}

\altaffiltext{*}{{\it Herschel} is an ESA space observatory with science instruments provided by European-led Principal Investigator consortia and with important participation from NASA.}
\altaffiltext{1}{ETH Zurich, Institute for Astronomy, Wolfgang-Pauli-strasse 27, 8093 Zurich, Switzerland}
\altaffiltext{2}{NASA Ames Research Center, Moffett Field, CA 94035, USA}
\altaffiltext{3}{SETI Institute, 189 Bernardo Ave., Mountain View, CA 94043, USA}
\altaffiltext{4}{University of Rochester, Department of Physics \& Astronomy, Rochester, NY 14627-0171, USA}
\altaffiltext{5}{Current address: Cerro-Tololo Inter-American Observatory, Casilla 603, La Serena, Chile}

\begin{abstract}
We present {\em Herschel} PACS spectroscopy of the\,[OI] 63\,\micron\ gas-line for three circumstellar disk systems showing signs of significant disk evolution and/or planet formation: HR 8799, HD 377 and RX J1852.3-3700. [OI] is undetected toward HR 8799 and HD 377 with $3\sigma$ upper limits of $6.8 \times 10^{-18}$ W m$^{-2}$ and $9.9 \times 10^{-18}$ W m$^{-2}$ respectively. We find an [OI] detection for RX J1852.3-3700 at $12.3 \pm 1.8 \times 10^{-18}$ W m$^{-2}$. We use thermo-chemical disk models to model the gas emission, using constraints on the [OI] 63 \micron\, and ancillary data to derive gas mass upper limits and constrain gas-to-dust  ratios. For HD 377 and HR 8799, we find 3\,$\sigma$ upper limits on the gas mass of 0.1--20 \mearth. For RX J1852.3-3700, we find two distinct disk scenarios that could explain the detection of [OI] 63 \micron\ and CO(2--1) upper limits reported from the literature: (i) a large disk with gas co-located with the dust (16--500 AU), resulting in a large tenuous disk with $\sim 16$ \mearth\ of gas, or (ii) an optically thick gas disk, truncated at $\sim 70$ AU, with a gas mass of 150 \mearth. We discuss the implications of these results for the formation and evolution of planets in these three systems.
\end{abstract}

\section{Introduction}
Protoplanetary disks are found around most, if not all, forming stars, with typical measured lifetimes of a few Myr \citep[e.g.][]{haisch01,meyer07}. Planet formation is thought to be ongoing in these first few Myrs, leading to gas giant planets, ice-giants, and planetesimal belts that interact with the gas. The interaction with the gas feeds the growth of the planets, causes their migration, and circularizes their orbits. A key constraint governing the formation of gas giants, ice-giants and terrestrial rocky planets, as well as their orbital evolution, is the timescale for gas clearing in a disk. Recently, the Kepler mission and radial velocity surveys have uncovered several candidate planets with sizes of order a few Earth that could be either rocky ``super Earths'' or gaseous ``mini Neptunes'' \citep[e.g.\ ][]{fressin12,gaidos12,croll11}

A sensitive search for the [SI] 25.23 \micron\ and molecular hydrogen gas emission lines in the infrared with {\em Spitzer} of 10--100 Myr old sun-like stars with dusty disks has shown that less than $\sim13$ -- 190 \mearth\  of gas at 100 K was present, even for the youngest stars in the sample with ages of 5--20 Myr \citep{pascucci06}, i.e. too little gas left to still form Jupiter--like gas giants. Furthermore, \citet{pascucci06} found CO derived gas limits of less than a few \mearth\ in the 10--40 AU region, for a sub-sample of 8 disks with ages of 10 to a few 100 Myr, excluding even the further formation of ice-giants like Uranus or Neptune in these systems. More data on a larger sample are warranted to address the question of how frequent these young 10--100 Myr stars with debris disk systems still have enough remnant gas (e.g. $\sim$10 \mearth) to form ice-giants like Uranus and Neptune. 

Work by \citet{gorti04,gorti08,hollenbach09} has demonstrated that the [OI] 63 $\mu$m emission is one of the strongest gas emission lines originating from the 10--50 AU regions of protoplanetary disks, and a sensitive probe for measuring small amounts of remnant gas, tracing the gas disk as it dissipates. Numerical simulations show that major collisions probably occurred in the ice-giant zone to 10 Myr or longer \citep{grigorieva07,ford07}, beyond the lifetime of optically thick disks. Such activity would release additional gas in either a pseudo-continuous or episodic fashion. Observations of [OI] are needed to trace small amounts of gas, and help address questions such as: how long does gas capable of forming ice-giants and circularizing the orbits of forming terrestrial planets persist? Is gas generation in this zone strongly variable from one system to another and hence possibly episodic? \citep{chen06,alexander07,besla07,kominami05}

The PACS instrument \citep{poglitsch10} on-board the {\em Herschel Space Observatory} \citep{pilbratt10} provides a major opportunity to probe the far-IR wavelength regime that covers strong gas lines such as the [OI] 63 \micron\ line. The Herschel GASPS program is conducting a large scale survey of Herbig Ae, T Tauri, and debris disk systems, and has reported [OI] 63 \micron\ gas detections toward several disk sources with ages of $\sim$ 4--12 Myr in \citet{mathews10,thi11,woitke11,tilling12}. 

In this paper, we focus on three well-characterized disk systems: HR 8799, HD 377, and RX J1852.3-3700. We describe our sample in Sect.\ 2, and the observations and data reductions in Sect.\ 3. In Sect.\ 4, we present thermo-chemical disk models and estimate gas mass (upper limits), and in Sect.\ 5, we discuss the implications for planet formation and evolution in these sources.

\section{Sample}
For our sample selection, we focused on 10--100 Myr old debris disk systems that were well characterized with ancillary observations and detailed dust disk models available, and that would be well complementary to the existing {\em Herschel} programs, in particular GASPS \citep[e.g., ][]{mathews10}. Each target either shows direct evidence of the presence of planets, or signs of significant evolution in the dust disk structure, such as asymmetries, gaps or holes, as derived from direct imaging and/or their SED. We summarize here our sample in stellar properties and signs of disk evolution / planet formation. 

\paragraph{HR 8799} is a nearby (d=39.4 $\pm$ 1.1 pc, \citealt{van-leeuwen07}) planetary system around an early-type star. We adopt the stellar properties derived by \citet{gray99}. It is noted that while the effective temperature is like that of an F0 star, it is observed to be a $\lambda$ Bootis type star, i.e.\ having anomalously weaker abundances in some of its metal lines, see \citet{gray99} and \citet{sadakane06} for a description of the full spectral classification. We adopt the discussion of the age and potential association with nearby stellar groups of HR 8799 described in \citet{hinz10} and therefore adopt the age range estimate of 60$^{+100}_{-30}$ Myr from \citet{marois08}. Four giant planets were discovered through direct imaging \citep{marois08,marois10}, with masses of 9, 10, 10, and 7 M$_{\mathrm{Jup}}$, located at projected separations of 14.5, 24, 38 and 68 AU respectively. \citet{close10} searched for the presence of new wide companions and concluded there are no companions with masses $>$3 \mjup\ within $\sim5''-15''$ ($\sim200-600$ AU). 

HR\,8799 has a remnant debris disk, first identified with IRAS photometry \citep{zuckerman04}. \citet{chen06} distinguished two components, a hot inner ring at 6 AU and a cooler outer disk from {\em Spitzer} observations. \citet{su09} revealed from {\em Spitzer} 70 \micron\ images there to be a large halo of small dust grains at large radii, and they presented a disk model to consist of: 1) warm (T$\sim$150K) dust at 6--15 AU, within the innermost planet; 2) a broad 90--300 AU zone of colder dust (T $\sim$ 45 K) beyond the outermost planet; and 3) a 300-1000 AU halo of small grains  probably being blown out of the system \citep{su09}. The disk was spatially resolved at 880 \micron\ observations by \citet{hughes11}, and found to be consistent with the presence of the 2nd component of the Su model, albeit with a larger inner radius of 150 AU (rather than 90 AU). \citet{patience11} resolved the cooler debris disk at 350 \micron, consistent with a 100--300 AU size, while also showing an indication of asymmetric structure. A dust mass for the debris disk of 0.1 \mearth\ has been derived based on sub-millimeter photometry \citep{williams06}.

\paragraph{HD 377} is a nearby (d=39.8 pc, \citealt{van-leeuwen07}) G2V star ($\sim$ 1.1 \msun) with an age estimate based on a variety of indicators, summarized in both \citet{apai08} and \citet{roccatagliata09}. We adopt an age of 150$^{+70}_{-125}$ (95\% CL) Myr.

HD 377 exhibits an SED consistent with a disk with a dust mass in small grains $> 4 \times 10^{-4}$ \mearth\ from $<$ 10 to $>$ 100 AU \citep{hillenbrand08} and $<$ 0.1 \mearth\ of warm gas (T $>$ 100K) within 2 AU \citep{pascucci06}. A detailed model including new sub--mm data \citep{roccatagliata09} indicates 0.06 \mearth\ in grains $<$ 3 mm located at orbital radii from 3--150 AU. It has not yet been resolved in scattered light. There is no evidence to date that it has formed a gas giant planet \citep{apai08}. 

\paragraph{RX J1852.3-3700} is a K3 (1.1 M$_{\odot}$) T Tauri star, in the general vicinity of the CrA star-forming region (d = 132 pc, \citealt{carpenter05}) - situated 1.84 deg away from the center of the CrA cluster, and co-moving with it in terms of both proper motion and radial velocity. We adopt photometric colors from \citet{neuhauser00}, and estimate an average Av of 0.89 $\pm$ 0.05 mag. We adopt a spectral type of K3 with +- 1 subtype and derive a $\log T = 3.682 \pm 0.020$, and $\log L = -0.226 \pm 0.087$, where L is in \lsun\ and T is in K units. Comparison with the \citet{baraffe98}, \citet{dantona97} and \citet{siess97} evolutionary tracks gives us an range in age estimates of $\sim10-23$ Myr. We adopt here an age of 18$^{+12}_{-7}$ Myr, derived from the Baraffe tracks.

RX J1852.3-3700 harbours a transitional disk with a low accretion rate \citep[$5 \times 10^{-10}$ \msun\ yr$^{-1}$, ][]{pascucci07}. In a comprehensive study utilizing the {\em Spitzer} SED, as well as sub--mm SMA and ASTE observations, and detailed models based on RADMC \citep{dullemond04a}, \citet{hughes10} inferred an optically--thin region within 16 AU and determined that the gas content in the system is about an order of magnitude lower than expected for the standard assumed gas-to-dust mass ratio of 100 and a CO abundance of $10^{-4}$ relative to H$_2$, suggesting gas clearing is likely underway in this system.

We summarize the key parameters of each source in Table \ref{table:starpar}.

\begin{deluxetable*}{lccc}
\tablecaption{Sample properties \label{table:starpar}}
\tablewidth{\linewidth}
\tablehead{
\colhead{ } & \colhead{HR 8799} & \colhead{HD 377} & \colhead{RX J1852.3-3700}
}
\startdata
Spectral Type & kA5 hF0 mA5 v $\lambda$ Boo\tablenotemark{a} & G2V\tablenotemark{b} & K3\tablenotemark{c} \\
Distance [pc] & 39.4$\pm$1.1\tablenotemark{d} & $39.1\pm1.1$\tablenotemark{d} & 130\tablenotemark{e} \\
Age [Myr] & 60$^{+100}_{-30}$\tablenotemark{f} & 150$^{+70}_{-125}$\tablenotemark{g} & 18$^{+12}_{-7}$\tablenotemark{g} \\
Stellar Mass [\msun] & 1.47\tablenotemark{a} & 1.02\tablenotemark{h} & 1.1\tablenotemark{c}\\
Stellar Temperature [K] & 7430\tablenotemark{a} & 5849\tablenotemark{h} & 4808\tablenotemark{g}\\
Stellar Luminosity [\lsun] & 4.92\tablenotemark{a} & 1.17\tablenotemark{h} & 0.59\tablenotemark{g}\\
Log(L$_{\mathrm{x}}$ / erg s$^{-1}$) [dex] & 28.11\tablenotemark{i}& 29.1\tablenotemark{j}& 30.41\tablenotemark{c}
\enddata
\tablerefs{a. \citet{gray99}, b. \citet{moore50}, c. \citet{neuhauser00}, d. \citet{van-leeuwen07}, e. \citet{carpenter05}, f. \citet{marois08}, g. this work, h. \citet{casagrande11}, i. \citet{robrade10}, j. \citet{wichmann03}}
\end{deluxetable*}

\section{Herschel observations}
\label{sec:observations}
Our targets were observed with {\em Herschel} PACS IFU spectrometer, with pointed chop/nod line spectroscopy AORs. The center spaxel ($9.4'' \times 9.4''$) of the $5 \times 5$ array was centered on the source in both nod positions and we utilized a small chop throw (2$''$). The observation centered on the [OI] $^3P_1\,\rightarrow\, ^3P_2$ transition at 63.1852 \micron, exposing for 2 line repetitions and 5 cycles, resulting in a nominal on-source integration time of 3840 sec. Observing date, program and AOR ID are listed in Table \ref{tab:observationlog}.
\begin{deluxetable}{llll}
\tablecaption{Observation log \label{tab:observationlog}}
\tablewidth{\linewidth}
\tablehead{
\colhead{Name} & \colhead{Date Observed} & \colhead{Program ID} & \colhead{Observation ID}
}
\startdata
HR 8799 & Jan. 1, 2011 & GT1\_vgeers\_1 & 1342212242\\
HD 377 & Jan. 11, 2011 & OT1\_vgeers\_2 & 1342212530\\
RX J1852.3-3700 & Mar. 15, 2011 & OT1\_vgeers\_2 & 1342216163
\enddata
\end{deluxetable}%

We reduced all PACS spectra with the Herschel Interactive Processing Environment \citep[HIPE, ][]{ott10}, v7.0.0, user release, build 7.0.1931, with PACS calibration tree version 26 \footnote{HIPE software obtained from \url{http://herschel.esac.esa.int/HIPE\_download.shtml}}. The spectra were extracted from the central spaxel, using the HIPE built-in correction for the point-source loss correction. 

At the time we carried out our spectral analysis, the latest HIPE version did not include formal propagation of the flux errors on individual pixels. Therefore, we estimated the noise by measuring directly the dispersion of the pixels from a first-order polynomial that was fitted to the continuum over short spectral ranges around the expected central wavelength of 63.1852 \micron\footnote{NIST Atomic Spectra Database v4.1.0 \citep{ralchenko11}} of the [OI] emission feature (63.03--63.16 \micron, and 63.22--63.39 \micron). This pixel dispersion was checked for Gaussianity with the Shapiro-Wilk test \citep{shapiro65} and found for all three sources to be consistent with being drawn from a normal distribution. The measured rms of the pixel dispersion is thus taken as a representative measure of the noise, and used further in our derivation of the line flux uncertainty and line flux upper limits. 

From the continuum fit, we report the continuum flux estimate or 3 $\sigma$ upper limits at 63.1852 \micron\ in Table \ref{tab:linefluxes}. We note the recommended PACS absolute calibration uncertainty for these observations is 30\% but do not include this in the uncertainties reported in Table \ref{tab:linefluxes}. We note that for our one source with a significant detection, RX J1852.3-3700, the estimated continuum flux at 63.19 \micron\ is consistent with the IRAS 60 \micron\ and Spitzer MIPS 70 \micron\ measurements to within $\sim 25\%$.

\begin{figure}[ht]
\begin{center}
\includegraphics[width=\columnwidth]{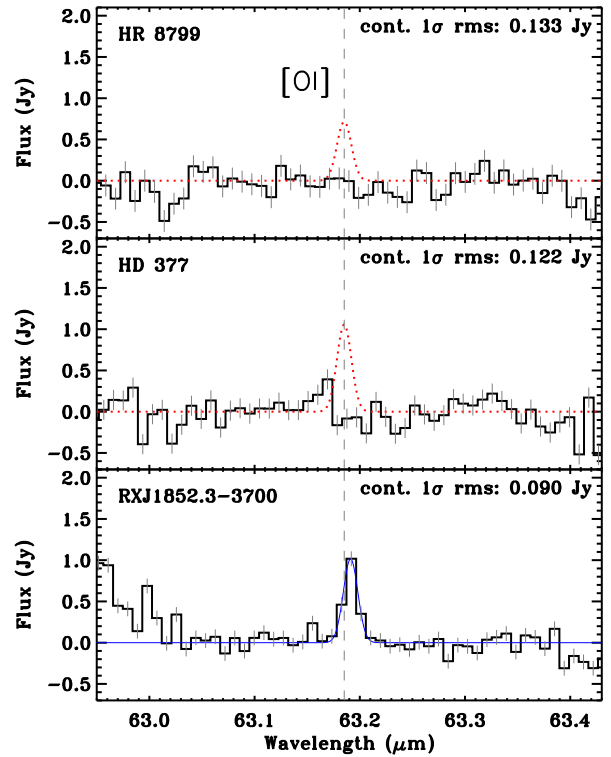}
\caption{{\em Herschel} PACS spectra for HR 8799, HD 377, and RX J1852.3-3700, with a linear continuum fit subtracted. The continuum fit is based on two short wavelength ranges on either side of the [OI] feature. The rest wavelength of the [OI] $^3P_1\,\rightarrow\, ^3P_2$ transition at 63.1852 \micron\ is indicated with a dashed line. The dotted lines indicate hypothetical [OI] feature with line flux equal to the 3 $\sigma$ upper limit and width equal to that derived for the RX J1852.3-3700 detection.}
\label{fig:specs}
\end{center}
\end{figure}

To estimate the integrated line intensity of the detected [OI] feature, we first perform a least-squares fit of a 1st order polynomial to the continuum, and then remove this baseline continuum fit. Next, we use the same approach to fit a Gaussian to the remaining spectrum, restricting the central wavelength to 63.18--63.20 \micron. Finally, the line flux is computed as the area of the best-fit Gaussian. The spectra, continuum and line fits are shown in Figure \ref{fig:specs}.

\begin{figure}[ht]
\begin{center}
\includegraphics[width=\columnwidth]{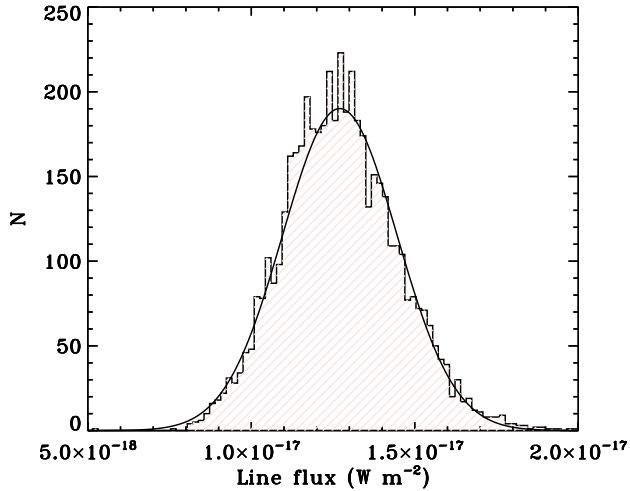}
\caption{Distribution of line fluxes derived 1000 times from ``resampled'' spectra for RX J1852.3-3700. The sigma of this distribution, derived with biweight\_mean, is used as the approximation of the uncertainty of the line flux.}
\label{fig:lfunc_rxj1852}
\end{center}
\end{figure}

The uncertainty on the line flux is evaluated using a Monte-Carlo approach. We start from the original spectrum, which we assume to represent the mean, and then we add pixel-by-pixel normally distributed random noise to the spectrum, where the amplitude of the random noise is set equal to the rms of the continuum flux dispersion estimated previously.  We then redo the line flux estimate on this new ``resampled'' spectrum, following the same method as described above. We repeat this procedure 5000 times and so create a distribution of estimated line fluxes for a distribution of resampled spectra. We take the robustly estimated dispersion of this distribution of line fluxes (using IDL's ``biweight\_mean'' routine) to be our formal uncertainty in the line flux. This method takes into account both the flux uncertainty in our spectrum (estimated from the rms) as well as the uncertainties on the fits to the continuum and the line. An example of this line flux distribution is shown in Figure \ref{fig:lfunc_rxj1852} for RX J1852.3-3700.

We consider the [OI] emission feature to be detected when the measured line flux is larger than 3 times the estimated uncertainty in the line flux. For sources with a non-detection, we estimate an upper limit on the line flux using the following approach. We start with the original spectrum, and add to it a fake [OI] feature in the form of a Gaussian, centered on 63.185 \micron, and with width equal to 0.0683 \micron, and we choose a specific amplitude taken from a predefined range of values. The value for width is taken from the best-fit to the source RX J1852.3-3700 for which [OI] is clearly detected. Based on the spectrum with fake [OI] feature thus created, we repeat the measurement of the line flux and its uncertainty as described above, and evaluate whether this fake feature would have been considered a detection. We repeat this latter for a range of increasing amplitudes for the fake feature, and search so for the minimum feature amplitude at which we would have ``detected'' the feature (3$\sigma$). We take the line flux corresponding to this minimum feature amplitude to be our upper limit on the [OI] line flux. 

The [OI] 63 \micron\ line is not detected toward HR 8799 and HD 377 in the central spaxel, nor in the other spaxels. The derived 3$\sigma$ upper limits are reported in Table \ref{tab:linefluxes}. 

The [OI] line is detected toward RX J1852.3-3700 with a line flux of $12.3 \times 10^{-17}$ W m$^{-2}$ and a $1\sigma$ uncertainty of $1.8 \times 10^{-17}$ W m$^{-2}$. We note here too this reported uncertainty does not include the absolute PACS calibration uncertainty of 30\%. The [OI] line appears spectrally unresolved, with a FWHM $\sim$ 77 km s$^{-1}$. Its line center wavelength of 63.191 \micron\ appears slightly redshifted to the rest wavelength, however, we note that this small shift is within 1$\sigma$ of the absolute wavelength calibration uncertainty\footnote{based on Fig.\ 12 and Sect.\ 5.2 of the PACS Spectroscopy Performance and Calibration document of May 31, 2011} at $\lambda = 63\ \micron$, due to nominal pointing uncertainty of order 2$''$. Inspection of the off-center IFU spaxels show no significant indication of extended [OI] emission. 

\begin{deluxetable*}{lllll}
\tablecaption{Summary of [OI] observations: observed central wavelength, line flux, continuum flux, and line luminosity. The [OI] and continuum detections are reported with 1\,$\sigma$ uncertainty, the non-detections are reported as 3\,$\sigma$ upper limits, and the reported uncertainties do not include the 30\% absolute PACS calibration uncertainty, see Sect.\ \ref{sec:observations}. \label{tab:linefluxes}}
\tablewidth{\linewidth}
\tablehead{
\colhead{Name} & \colhead{$\lambda$} & \colhead{Line Flux} & \colhead{Cont. Flux} & \colhead{Line Luminosity} \\
\colhead{ } & [\micron] & \colhead{[$10^{-18}$ W m$^{-2}$]} & \colhead{[Jy]} & \colhead{[$10^{-6}$ \lsun]}
 }
\startdata
HR 8799 & - & $<$ 6.8 & $<$ 0.40 & $<$ 0.32\\ %checked 
HD 377 & - & $<$ 9.9 & $<$ 0.37 & $<$ 0.47\\  % checked
RX J1852.3-3700 & 63.191 & 12.3 $\pm$ 1.8 & 1.82 $\pm$ 0.09 & 6.7 $\pm$ 0.4
\enddata
\end{deluxetable*}

%%%
\section{Modeling}

\subsection{Model description}
We calculate gas line emission using detailed thermo-chemical disk models as described in \citet{gorti04,gorti08,gorti11} and \citet{hollenbach09}. We summarize the main features of the models here for completeness. Disk structure models are computed from an input surface density distribution ($\Sigma(r)$, where $r$ is the disk radius) by assuming vertical hydrostatic equilibrium.   Vertical gas density and temperature structure is obtained by thermal balance. The main heating mechanisms considered are heating due to collisions with warmer dust grains, grain photoelectric heating by Far Ultraviolet (FUV; 6\,eV$<h\nu<13.6$\,eV) photons,  X-rays, exothermic chemical reactions and cosmic rays. Gas cools by collisions with cooler dust grains and line emission by various ions, atoms and molecules whose abundances are calculated simultaneously by solving for disk chemistry. Our chemical network consists of 85 species and $\sim 600$ reactions that include photoionization and photodissociation processes. Freezing on grains to form ices is not included explicitly, but its effect on line emission is estimated when the dust temperatures are low enough for freeze-out. Gas radiative transfer is calculated by using an escape probability formalism and a simple two-layer model is used for the dust radiative transfer. 

\subsection{Input parameters}
The stellar spectrum (optical to X-ray wavelengths) is a primary input to the model and obtained from the literature for the disks in this study. Dust parameters such as the size distribution, composition, mass and radial extent are constrained by fitting the dust spectral energy distribution (SED). Extensive prior modeling of dust SEDs has been carried out for all the three disks considered, and we avail ourselves of these results. As the vertical structure in our models is determined by the {\em gas} density and temperature, we verify the fit to the dust SED after arriving at a disk solution. The abundance of PAHs, which affect gas temperature via grain photoelectric heating \citep[e.g.,][]{bakes94}, is unknown and we have set this to zero. For the disks around HR 8799 and HD 377, the dust grain size distribution inferred from SED modeling \citep{su09,roccatagliata09} shows a lack of sub-micron sized grains and this is consistent with an insignificant PAH population in these disks. However, the disk around RX J1852.3-3700 may have a more primordial dust grain size distribution and PAHs may be present. We adopt interstellar cloud gas phase abundances from \citet{savage96}. The adopted stellar and dust properties are listed in Tables~\ref{table:starpar} and \ref{table:dustpar} respectively.

\begin{deluxetable*}{lccccl}
\tablecaption{Adopted Dust Parameters \label{table:dustpar}}
\tablewidth{\linewidth}
\tablehead{
\colhead{Source} & \colhead{Surface Density} & \colhead{Radial Extent} & \colhead{Dust size} & \colhead{Dust Mass} & \colhead{Reference} \\
\colhead{ } & \colhead{$\Sigma(r) \propto$} & \colhead{[AU]} & \colhead{[\micron]} & \colhead{[\mearth]} & \colhead{ }
}
\startdata
HR 8799 \\
\hspace{0.5cm}{\it Inner Disk} & $r^0$ & $6<r<15$ & $1.5<a<4.5$ & 1.1E-6 & \citet{su09}\\
\hspace{0.5cm}{\it Outer Disk}& $r^0$ & $90<r<300$ & $10<a<1000$ & $0.12$ & \citet{su09}\\
HD 377& $r^0$ & $6.1<r<150$ & $14<a<3000$ & $0.058$ & \citet{roccatagliata09}\\
RX J1852.3-3700 & $\left(\frac{180\,\mathrm{AU}}{r}\right) e^{-r/180\,\mathrm{AU}}$ & $16<r<500$ & $0.1<a<1000$ & $53.3$ & \citet{hughes10}
\enddata
\end{deluxetable*}

\subsection{Model Results: Gas Content of Disks}

\paragraph{HR 8799} 
We consider that gas may be spatially co-located with the dust, either in the inner disk ($6<r<15$\,AU) or in the outer disk ($90<r<300$\,AU) and derive upper limits to the gas mass.  Although there could be gas present in the $15-90$\,AU region, this is dynamically unlikely due to the presence of multiple planets at these radii. We separately constrain gas mass in the inner and outer disks, i.e., we assume that gas is present in either one of these regions. We compute the expected [OI]\,63\,$\mu$m line luminosity for disks by varying gas mass and thus derive upper limits based on the PACS non-detection for this source. 

We find that the gas mass has to be lower than $\sim 0.1$\,\mearth\ in the inner disk and $\lesssim 5-8$\,\mearth\ in the outer disk in order to agree with the [OI] flux limits. For the inner disk model that corresponds to this upper limit, the densities and temperatures in the [OI]-emitting regions of the disk are $n\sim10^8$ cm$^{-3}$ and $T_{\mathrm{gas}}\sim 100-300$\,K. Heat released by photoreactions, mainly ionization of carbon, dominates gas heating. X-ray and FUV fluxes from the star are low and the absence of PAHs and low dust collisional cross-section per H atom lead to low efficiency in heating the gas. [OI] is the main coolant. In the outer disk model corresponding to the upper limit, the disk is extended and cold, with the gas temperature in the [OI] emitting region $T_{\mathrm{gas}} \sim 40-70$\,K, and gas densities  $n \sim 10^6$\,cm$^{-3}$. Photoionization of carbon again dominates gas heating and [OI]\,63\,$\mu$m and [CII]\,158\,$\mu$m lines are strong coolants. The gas/dust ratio upper limit in the outer disk is consistent with the canonical interstellar value of 100, while we find no useful constraint on the inner disk ($<10^5$).

\paragraph{HD 377} 
In addition to the limit on [OI] emission presented in this paper, this source has flux limits on H$_2$ S(1) emission and CO($2-1$) and CO($3-2$) rotational line emission of $2.2\times 10^{-18}$, $5\times10^{-20}$ and $2.2\times10^{-19}$\,W m$^{-2}$ respectively \citep{pascucci06}. We derive an upper limit on the gas mass between $6-150$\,AU of $\sim 1\,$\mearth\ based on all these constraints, and a corresponding gas/dust ratio limit of $< 17$. We note that the constraint derived from [OI] is more stringent than the limit of $\lesssim 20\,$\mearth\ for T=100 K gas based on H$_2$ flux limits \citep{pascucci06}.

\paragraph{RX J1852.3-3700}
The PACS [OI]\,63\,$\mu$m detection, {\em Spitzer} IRS [NeII]\,12.8\,$\mu$m detection, and the non-detection of both H$_2$ S(1)  and CO($2-1$) emission are all used for comparison with model calculated line fluxes to determine the distribution of gas in the disk. The fluxes and upper limits for these lines are $1.2\times10^{-17}$\,W m$^{-2}$ for [OI] (this paper),   $4.3\times10^{-18}$\,W m$^{-2}$ for [NeII] \citep{pascucci06}, $< 1.0\times10^{-18}$\,W m$^{-2}$ for H$_2$ S(1) \citep{pascucci06} and  $<  8\times10^{-22}$\,W m$^{-2}$ for CO($2-1$) \citep{hughes10}.

First, we consider constraints set by [OI] and CO emission and models where the inner hole region ($r<16$\,AU) is devoid of gas. [OI] and CO($2-1$) emission arises from relatively cool gas and these lines are better at tracing the disk mass.  There are then two possible disk configurations that can give rise to the observed [OI] and are still consistent with the non-detection of CO.

In one case, the gas disk is extended and spatially co-located with the dust (up to 500\,AU) and the mass of gas is very low, with the [OI] emission being optically thin. In such low mass, tenuous disks, CO is photodissociated by stellar and interstellar UV photons and there is no detectable CO($2-1$) emission. We derive a gas mass of  $\sim 16$\,\mearth\  within $16-500$\,AU in this case, and a highly reduced gas/dust mass ratio $\lesssim 1$.  We note that although this model disk extends to $500$\,AU, both the CO($2-1$) and [OI] emission decline sharply beyond $r\sim 200$\,AU and we do not probe gas in the $200-500$\,AU regions. Higher disk masses are possible, if we consider the possibility that CO might be frozen out on cold dust grains. In models that include the effects of photodesorption, and assume freezing of CO and of O as water ice when the dust temperature is below some critical value, we estimate slightly higher disk masses. We assume that CO and H$_2$O ice freeze at dust grain temperatures of 20 K and 80 K respectively, and account for photodesorption of ices as described in \citet{gorti11}. With these assumptions, we find a maximum disk mass of $\sim 40$ \mearth, and a gas/dust mass ratio of $\sim 3$ in the disk extending from 16--500 AU.

An alternate model is a truncated gas disk with the fiducial gas/dust mass ratio of 100 and is optically thick. This model disk produces excess [OI] and CO($2-1$) emission if it is as large as the dust disk and has to be limited to a radial extent of  $\sim 70$\,AU to explain the observations. The gas mass for the best fit model is $150$\,\mearth\ from $16-70$\,AU. Freezing effects are not significant for the warmer dust temperatures of the compact disk configuration. There is also a possibility that CO might be under-abundant in the disk, as is often inferred for other disks \citep[e.g.,][]{qi06,qi11}. In such a case, we infer CO depletion factors of $\sim 30-100$ in the disk around RX J1852.3-3700. We next include the constraints introduced by mid-infrared line emission.  The non-detection of H$_2$ S(1) excludes the possibility of a ``full" gas disk, i.e., a disk of mass $0.016$\,M$_{\odot}$ ($5.3 \times 10^3$ \mearth) with a gas/dust ratio of 100. All disk configurations with masses a factor of 10 or more lower are possible within the constraints set by H$_2$ S(1), corresponding to upper limits to the gas/dust ratio of $\lesssim 10$, consistent with the findings of \citet{hughes10}.  [NeII] emission is not a sensitive tracer of disk mass, and is reproduced to within a factor of $\sim 3$ in all our disk models. Model calculations always give lower [NeII] emission than observed in the models that assume a 16\,AU hole, which suggests that trace amounts of gas may be present within the hole. \citet{pascucci07} estimate an accretion rate of $5\times10^{-10}$\,M$_{\odot}$\,yr$^{-1}$ from the H$\alpha$ equivalent width, indicating the presence of at least some gas at $r<16$\,AU.

We calculate disk life times for the models using our time-dependent photoevaporation models \citep{gorti09a,gorti09b}. For the dense, truncated disk extending from $16-70$\,AU we find a disk lifetime of $\sim$0.5 Myr, while for extended, tenuous low mass disk we find a very short disk lifetime of only $\sim 4 \times 10^4$ yr. This argues in favor of the dense truncated disk model as the more likely scenario.

\section{Discussion}
Here we explore some of the implications of our findings. We first compare our [OI] detections and upper limits with what has so far been found for the GASPS sample. Then, we explore scenarios for potential further planet formation in our target sample.

\subsection{Comparison with the GASPS program}
The Herschel GASPS program is carrying out a gas line emission survey toward a sample of 240 young stars, focussing primarily on [OI] 63.185 \micron, [OI] 145.525 \micron\, and [CII] 157.741 \micron. GASPS have presented [OI] 63 \micron\ gas observations toward 7 disk sources with ages of $\sim$ 4--12 Myr in \citet{mathews10,thi11,woitke11,tilling12}. [OI] 63 \micron\ has been detected toward 6 of these 7 sources, with 1 non-detection for HD 181327. For the [OI] 63 detections in GASPS, the reported lines fluxes are typically a factor of a few stronger than what we find for the $\sim18$ Myr old T Tauri disk RX J1852.3-3700 (12.3 $\times10^{-18}$ W m$^{-2}$). Of particular note is the 4 Myr old Herbig Ae star HD 163296, which displays a very strong [OI] 63 \micron\ line flux of $\sim193 \times10^{-18}$ W m$^{-2}$. The [OI] 63 \micron\ upper limits that we find for our two older (60--150 Myr) disk sources are comparable to the $< 9.3 \times10^{-18}$ W m$^{-2}$ presented for HD 181327 (12 Myr old F5 star at 50.6 pc) \citep{mathews10}. 

[OI] 63 \micron\ is detected in one of the transitional disks in the GASPS sample, TW Hya, a 10 Myr K6V star with an inner 4 AU cavity in its disk. \citet{thi10} find that while the disk is still massive (0.5--5 $\times$ 10$^{-3}$\msun), a significant fraction of the primordial gas has already disappeared (gas-to-dust ratio between 2 and 26), possibly due to evaporation by the strong X-ray flux from TW Hya \citep[cf.\ ][]{pascucci09}.

For HD 169142, a $\sim6$ Myr Herbig Ae disk, \citet{meeus10} also detect [OI] 63 \micron\ gas, and they propose this disk may be transitional. With a possible 5--20 AU hole, and a reduced gas-to-dust ratio between 22 and 50, the disk is still quite gas rich with a gas mass of 3--7 $\times$ 10$^{-3}$\msun. 

We note that the results from our [OI] 63 \micron\ observations are overall consistent with those reported from the GASPS survey so far. More data are needed to derive a significant empirical relation for the [OI] 63 \micron\ line strength as a function of age and stellar mass. 

\subsection{Implications for planet formation}
With strong limits on the remnant gas in our sample of 3 evolved circumstellar disks, we can explore the possibilities for on-going planet formation. To address this, we estimate the size that we could expect proto-planets to grow through collisional accretion (``isolation mass'') given the ages of these systems, at the radii that a debris disk is still observed to be present. To estimate the isolation mass as a function of orbital radius, we use the standard prescription by \citet{ida08}: 
\begin{equation}
\small
M_{c,\mathrm{iso}} \simeq 0.16 \eta_{\mathrm{ice}}^{3/2} f_d^{3/2}\left(\frac{a}{1 \mathrm{AU}}\right)^{3/4-(3/2)(q_d-3/2)} \left(\frac{M_*}{M_{\odot}}\right)^{-1/2} M_{\oplus},
\end{equation}
where we assume $q_d = -1$ for  the slope of the power law for the dust mass surface density $\Sigma_d$, where $f_d$ is a multiplicative factor used to scale $\Sigma_d$, and $\eta_{\mathrm{ice}}$ is a step function describing the change in dust surface mass density at the ice-line. From this, the isolation mass scales with orbital radius and stellar mass as
\begin{equation}
M_{c,\mathrm{iso}} \propto a^{4.5} M_*^{-1/2}
\end{equation}
Next, we estimate what is the timescale for growing a planet up to this isolation mass through growth of solids, for given orbital radius and stellar mass, with the relation
\begin{equation}
t \sim \frac{\rho_p R_p}{\Sigma_d \Omega_p}
\end{equation}
from \citet{goldreich04}\footnote{this approximation does not include gravitational focussing, which means these are rather upper limits on the timescale}, with planet density $\rho_p$, planet radius $R_p$, surface mass density $\Sigma_d$, and planet orbital frequency $\Omega_p$. With $\Omega_p \propto a^{-3/2} M_*^{1/2}$, assuming $\Sigma_d \propto a^{-1} M_*$, and setting the planet mass to the isolation mass, it follows that this timescale scales with orbital radius and stellar mass as:
\begin{equation}
t \propto a^{4}\ M_*^{-5/3}
\end{equation}
We normalize these relations to the mass surface density of a disk that would be capable of forming an $\sim$1 \mearth\ isolation mass core at 5 AU (beyond the ice line) in 1 Myr around an 1 \msun\ star, and then scale these relations to the different stellar masses and disk radii appropriate for our sample. 

Following these approximate estimates for isolation mass and formation timescales, and given the ages of our sample disks, we find the following:

For {\em HD 377}, at an age of 150 Myr, several cores of a few earth mass could have formed already within the 6--150 AU debris disk. Our gas mass upper limit of $\lesssim$ 1 \mearth\ suggests there is not enough gas mass left to still form any more ice giants, but our upper limits cannot rule out the possibility that some existing gas / ice giants or dwarf planets could still be accreting remnant gas. The H$_2$ and He might accrete as a thin atmosphere, while the ``metal'' content ($Z>2$) would be expected to freeze out as solids and/or form an icy surface layer.

In {\em HR 8799}, with a derived gas mass upper limit of $\lesssim 5-8$ \mearth\ in the outer disk (90--300 AU), there is not enough gas still present beyond 90 AU for the further in-situ formation of gas giants. At the age of 60 Myr, only small dwarf protoplanets (of order 10 times smaller than Pluto size) could be expected to have grown through collisional accretion in the debris disk beyond a radius of 90 AU. With the upper limit of 5--8 \mearth, such small protoplanets would accrete some of this material to form a solid icy surface layer. 

In the {\em RX J1852.3-3700} system, we have distinguished two possible disk scenarios that could be consistent with the detected [OI] emission. In the dense truncated disk scenario, with an estimated gas mass of $\sim$ 150 \mearth\ and gas/dust mass ratio of $\sim100$, there could still be enough gas present to form icy giants or volatile rich super-earths beyond 16 AU. The isolation mass beyond 16 AU is estimated to be of order a hundred Earth masses and this could be reached on timescales of $\sim100$ Myr. At the estimated age of $\sim$18 Myr, a few \mearth\ core may have formed beyond 16 AU, for which there is probably enough gas left to yield a final gas to dust ratio of 1:1 as inferred for the ice giants in our solar system. In this dense disk scenario, the inner hole may have been carved out when the disk was still quite massive (through planet formation and/or X-ray photo-evaporation, e.g.\ \citealt{marsh92,owen11}), while the outer disk was ablated by FUV photo-evaporation \citep{gorti09a}. The gas would have been present in the giant and ice giant planet formation regions of a putative planetary system, even at very late stages, affecting gas content and dynamics. The gas disk would have dispersed by erosion at both the inner and outer edges and evolve into a Òtorus-likeÓ structure like we see at present. 
In the alternative scenario, we considered an extended gas disk co-located with the dust out to 500 AU with a much lower gas mass of 16 \mearth\ and highly reduced gas/dust mass ratio ($\lesssim$ 1). In such a disk, there would be too little gas left for the further formation of Jovian-type planets, although potentially enough to still influence the dynamics of already formed planets. The surface mass density of the disk could have declined through photo-evaporation \citep[e.g., by EUV photons,][]{alexander06b}, and viscous spreading, growing the disk to large radii. 

For future investigation, it will be interesting to explore the effect of gas drag on the particle growth as well as the damping of orbital eccentricities in these tenuous outer disks, and to consider global scenarios for planetary architecture that take into account the self-consistent formation and evolution (i.e.\ migration) scenarios as described in e.g.\ \citet{walsh11}. 

\section{Conclusions}
We present {\em Herschel} observations of the [OI] 63 \micron\ gas line for three older debris disks, HR 8799, HD 377, and RX J1852.3-3700. [OI] is undetected toward HR 8799 and HD 377 with $3\sigma$ upper limits of $6.8 \times 10^{-18}$ W m$^{-2}$ and $9.9 \times 10^{-18}$ W m$^{-2}$ respectively, while detected toward RX J1852.3-3700 with a line flux of $12.3 \pm 1.8 \times 10^{-18}$ W m$^{-2}$. We use thermo-chemical disk models to model the gas emission, using constraints on the [OI] 63 \micron\, and available ancillary data on the dust structure, to derive gas mass upper limits and constrain gas-to-dust ratios. Based on the non-detections for HR 8799 ($60^{+100}_{-30}$ Myr) and HD 377 ($150^{+70}_{-125}$ Myr), we derive gas mass upper limits between 0.1--1.5 \mearth\ for these sources. For the [OI] detection toward RX J1852.3-3700 (18 Myr), we infer two possible scenarios for either a smaller $16-70$\,AU optically thick disk of $150$\,\mearth\, or alternatively a large $16-500$\,AU optically thin disk with about $\sim 16$\,\mearth\ gas mass that has a significantly reduced gas/dust ratio of $\sim$1. It appears that there is no possibility for further ice giant formation for HD 377 and HR 8799,  which could only still be forming smaller dwarf planets. In RX J1852.3-3700, there appears to be sufficient solid and gaseous material for an ice giant to still form beyond 16 AU.

\section*{Acknowledgements}
U.G. and D.H. acknowledge support from grants under the NASA Herschel Data Analysis program and the NASA Astrophysics Data Analysis Program (NNX09AC78G) which enabled this work. E.M. acknowledges support from JPL contract 1433135. This work is based on observations made with {\it Herschel}, a European Space Agency Cornerstone Mission with significant participation by NASA. Part of the observations were acquired in time guaranteed by a HIFI hardware contribution funded by Swiss PRODEX (grant 13911/99/NL/SFe). HIFI has been designed and built by a consortium of institutes and university departments from across Europe, Canada and the United States under the leadership of SRON Netherlands Institute for Space Research, Groningen, The Netherlands and with major contributions from Germany, France and the US. Consortium members are: Canada: CSA, U.Waterloo; France: CESR, LAB, LERMA, IRAM; Germany: KOSMA, MPIfR, MPS; Ireland, NUI Maynooth; Italy: ASI, IFSI-INAF, Osservatorio Astrofisico di Arcetri-INAF; Netherlands: SRON, TUD; Poland: CAMK, CBK; Spain: Observatorio Astron\'{o}mico Nacional (IGN), Centro de Astrobiolog\'{\i}a (CSIC-INTA). Sweden: Chalmers University of Technology - MC2, RSS \& GARD; Onsala Space Observatory; Swedish National Space Board, Stockholm University - Stockholm Observatory; Switzerland: ETH Zurich, FHNW; USA: Caltech, JPL, NHSC. HIPE is a joint development by the Herschel Science Ground Segment Consortium, consisting of ESA, the NASA Herschel Science Center, and the HIFI, PACS and SPIRE consortia.

\bibliographystyle{apj}

\end{document}